# Mapping the structural, magnetic and electronic behaviour of $(Eu_{1-x}Ca_x)_2Ir_2O_7$ across a metal-insulator transition

Eli Zoghlin[1], Zach Porter[1], Samuel Britner[2], Samra Husremovic[2], Yongseong Choi[3], Daniel Haskel[3], Geneva Laurita[2] and Stephen D Wilson[1]

[1] Materials Department, University of California, Santa Barbara, California 93106, USA
[2] Department of Chemistry and Biochemistry, Bates College, Lewiston, Maine 04240, USA
[3] Advanced Photon Source, Argonne National Laboratory, Argonne, Illinois 60439, USA

E-mail: stephendwilson@ucsb.edu

## Abstract

In this study, we employ bulk electronic properties characterization and x-ray scattering/spectroscopy techniques to map the structural, magnetic and electronic properties of $(Eu_{1-x}Ca_x)_2Ir_2O_7$ as a function of Ca-doping. As expected, the metal-insulator transition temperature, $T_{MIT}$, decreases with Ca-doping until a metallic state is realized down to 2 K. In contrast, $T_{AFM}$ becomes decoupled from the MIT and (likely short-range) AFM order persists into the metallic regime. This decoupling is understood as a result of the onset of an electronically phase separated state, the occurrence of which seemingly depends on both synthesis method and rare earth site magnetism. PDF analysis suggests that electronic phase separation occurs without accompanying chemical phase segregation or changes in the short-range crystallographic symmetry while synchrotron x-ray diffraction confirms that there is no change in the long-range crystallographic symmetry. X-ray absorption measurements confirm the $J_{eff} = ½$ character of $(Eu_{1-x}Ca_x)_2Ir_2O_7$. Surprisingly these measurements also indicate a net electron doping, rather than the expected hole doping, indicating a compensatory mechanism. Lastly, XMCD measurements show a weak Ir magnetic polarization that is largely unaffected by Ca-doping.

## 1. Introduction

The highly accommodating nature of the pyrochlore structure allows for decoration of a geometrically frustrated lattice with a wide array of different chemistries, leading to a myriad of interesting behaviours. In particular, the pyrochlore iridates ($A_2B_2O_7$, where A = Y or a lanthanide and B = Ir) have attracted significant attention due to their comparable energy scales for spin-orbit coupling (SOC) and electron-electron correlations $(U)$ which lead to $J_{eff} = ½$ Mott physics [1]. In these compounds, a thermally driven metal-to-insulator transition (MIT) occurs for A = Nd and higher atomic number (Z), with a gap between quadratic bands opening at $T_{MIT}$. $T_{MIT}$ decreases in magnitude with increasing lanthanide ionic radius until $Pr_2Ir_2O_7$ is reached and a metallic ground state is formed, driven by the change in Ir valence band bandwidth [2]. The MIT is accompanied by an AFM transition ($T_{AFM}$) with the formation of the $q = 0$, 'all-in-all-out' (AIAO) magnetic order on the Ir and lanthanide sublattices (see **figure 1**) [3, 4]. This magnetic order is a result of the tetrahedral arrangement of the Ir $J_{eff} = ½$ spins, which frustrates the (antiferromagnetic) Heisenberg exchange interaction. The frustration is lifted by the action of the Dzyaloshinsky-Moriya interaction (DMI), amplified by the strong

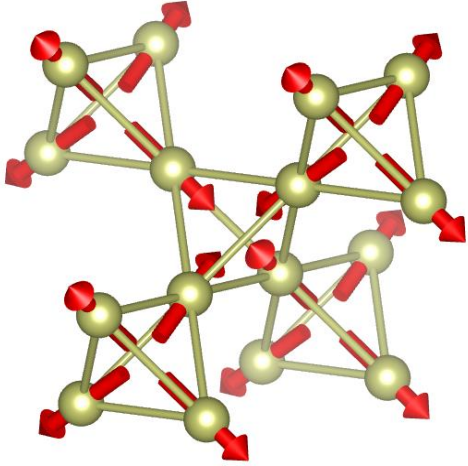

**Figure 1:** The "all-in-all-out" (AIAO) magnetic structure on the Ir sublattice in $Eu_2Ir_2O_7$. Only the Ir atoms are shown

SOC. Since the Heisenberg exchange is frustrated, the (weaker) DMI determines the magnetic order, resulting in the AIAO state [5, 6, 7].

Based on these ingredients, theoretical predictions have suggested that suppression of the MIT can lead to a novel antiferromagnetic quantum critical point [8] and the stabilization of topologically nontrivial phases [5, 9, 10]. Signatures of the Weyl semimetal phase have been observed immediately below $T_{MIT}$, in addition to other atypical electronic behaviours [11, 12, 13, 14, 15]. Suppression of the MIT through modification of the ratio $U/W$ ($W$, valence band bandwidth) has been attained by both physical [16, 17] and chemical pressure (i.e. alloying with different lanthanide ions) [17]. While these experimental techniques provide slightly different structural perturbations, both represent methods of bandwidth-control of the MIT.

Despite these efforts, no clear consensus has emerged about the mechanism for the MIT in this system and its relationship to the Ir AIAO order. Understanding this connection is vital for evaluating the nature of the metallic state formed and the potential for quantum critical phase behaviour. $Eu_2Ir_2O_7$ provides a convenient platform for this investigation since the nonmagnetic nature ($|L| = |S|$, $J = 0$) of the $Eu^{3+}$ ion allows for direct study of the Ir AIAO magnetism absent the influence of lanthanide order. Illustratively, ARPES work on $Nd_2Ir_2O_7$ shows a continuous gap evolution immediately below $T_{MIT}$, consistent with a mean-field mechanism that directly connects the onset of the AIAO order with the insulating behaviour [14]. However, a more Mott-like state emerges at lower temperatures, possibly due to the ordering of the Nd moments [14]. Detailed magneto-resistance studies on $Nd_2Ir_2O_7$ samples further show a complex interplay of the Nd/Ir magnetic orders with the charge transport due to the presence of $f - d$ exchange coupling [12, 18], furthering the motivation to study compounds with a nonmagnetic A-site ion.

In contrast to the bandwidth-control studies, heterovalent doping of the A-site ion allows for a detailed survey of the changes in the charge transport and structural/magnetic symmetries as the valence band deviates from half-filling. Here, Ca-doping is used for this purpose since it provides a stable 2+ ion with a minimal difference in ionic radius – for comparison, $Sr^{2+}$ is ≈ 20% larger than $Eu^{3+}$ while $Ca^{2+}$ is only 5% larger. Ca-doping therefore allows for modification of the Ir 5d occupation (i.e. filling-control) and the SOC-driven AIAO state – affording an opportunity to study the connection to the MIT – with minimal structural perturbation. Traversing the iridate phase diagram via filling-control may reveal significantly different magnetic and electronic states than what has been observed through bandwidth-control [8]. However, reports in the literature of carrier-doping iridates show varied results. With respect to the relationship between $T_{MIT}$ and $T_{AFM}$, these variations can be summarized in three distinct cases: (1) $T_{MIT} = T_{AFM}$ is maintained with doping until a fully metallic/paramagnetic state is realized [19] (2) Both $T_{AFM}$ and $T_{MIT}$ decrease with doping, but $T_{MIT}$ decreases much faster, becoming decoupled from $T_{AFM}$ [20] (3) A second, higher temperature, feature in the magnetic susceptibility emerges upon doping and eventually dominates the signal; $T_{MIT}$ still decreases, but the original $T_{AFM}$ appears essentially unchanged [21]. Addressing how these different cases relate to the synthesis procedure and resulting sample composition is important in order to determine the intrinsic properties of the pyrochlore iridates.

In this study, we utilize carrier-doping to drive a change in the ground state of $Eu_2Ir_2O_7$. We leverage bulk characterization and x-ray scattering/spectroscopy to detail the changes that occur in the crystallographic, magnetic and electronic structures across the series ($Eu_{1-x}Ca_x)_2Ir_2O_7$. $T_{MIT}$ is rapidly suppressed with doping until a metallic ground state is reached between $.04 < x < .07$. In parallel, $T_{AFM}$ decreases slowly and becomes decoupled from $T_{MIT}$ such that the magnetic order of the parent compound survives well into the metallic regime. This behaviour is interpreted as being due to the formation of an electronically phase separated state. An open question remains as to why this phase separation is seen only under certain synthetic conditions and not for samples with a magnetic A-site ion. X-ray scattering measurements indicate that the modification of $T_{MIT}$ and $T_{AFM}$ is not coincident with observable changes in the average or local structural symmetries. Analysis of Ir $L$ edge x-ray absorption spectroscopy (XAS) shows a clear net change in valence with doping, confirming the occurrence of filling-control. Unexpectedly, this filling-control manifests as electron doping, rather than the expected hole doping, likely due to a compensatory defect mechanism.

**2. Methods**

Polycrystalline samples of $(Eu_{1-x}Ca_x)_2Ir_2O_7$ were made by a solid-state synthesis procedure. Stoichiometric mixtures of dry $Eu_2O_3$ (Alfa Aesar, 99.99%), $IrO_2$ (Alfa Aesar, 99.99%),) and $CaCO_3$ (Alfa Aesar, 99.99%) powders were ground together until well mixed. The loose powder was then placed in an $Al_2O_3$ crucible and sintered in air at 1073 K for 18 hours. The powder was then re-ground, formed into a dense pellet using a cold-isostatic press at 350 MPa and fired in air at 1273 K for 100 hours. In the initial samples used for the x-ray scattering/spectroscopy studies, this process was repeated at 1373 K following re-grinding and re-pressing [22]. These samples are referred to as "air sintered" in the text. Later samples were synthesized following a modified procedure utilizing lower temperatures in air and a vacuum sintering step to reduce $IrO_2$ volatility [23]. Following the same initial 1073 K treatment, pellets were sintered at 1298 K three times for 100 hours with intermediate re-grinding and re-pelletizing. Next, 2 – 4 molar percent of $IrO_2$ was added to the powder before sintering at 1298 K for 100 hours in an $Al_2O_3$ crucible sealed under vacuum in a quartz tube. In some cases, this step was repeated for more complete reaction of the $Eu_2O_3$ precursor. Finally, after regrinding and repressing, the pellet was fired at 1173 K for 48 hours in air. These samples are referred to as "vacuum sintered".

Evaluation of the final doping level ($x$) was performed using a Rigaku ZSX Primus IV wavelength-dispersive x-ray fluorescence (WDXRF) spectrometer. Quantitative analysis was attained by use of a standard curve created by measurement of pressed pellets of unreacted precursor powders with known stoichiometry. A range of dopings of the air-sintered samples were studied by high-resolution, synchrotron powder x-ray diffraction (XRD) at the 11-BM beamline of the Advanced Photon Source (APS). In order to minimize the effect of absorption, a thin layer of finely ground powder was adhered using grease to the outside of a nested Kapton tube. Laboratory XRD using a Panalytical Emperyan diffractometer was utilized to monitor the progression of the reaction during the synthesis procedure and evaluate structural changes in the vacuum sintered samples. All diffraction data were refined using the TOPAS software package. VESTA was used to visualize the results of the refinements and determine the bond angles and distances [24]. The pyrochlore structure was found to be the majority phase in all samples measured at 11-BM ($\approx$ 95% purity). Impurities consisted of paramagnetic $IrO_2$, Ir metal and $Eu_2O_3$. $IrO_2$ and Ir were found at levels typically ≤ 1.5% while the fraction of unreacted $Eu_2O_3$ was generally larger, up to 2.5% in some samples. The vacuum sintered samples were found to have a higher $Eu_2Ir_2O_7$ phase fraction due to a more complete reaction of the $Eu_2O_3$. Based on lab XRD these samples contained smaller amounts of $IrO_2$, Ir and $Eu_2O_3$ (≈ 1% each).

Synchrotron total scattering data for pair distribution function (PDF) analysis were collected on the 6-ID-D beamline at the APS using powder taken from the same batches as the 11-BM samples. The well-ground powder was sieved using 325 mesh wire cloth (.044 mm opening size) to prevent pin-holing. This powder was then sealed into Kapton tubes using copper wire and epoxy in a He-filled glove-bag to provide a thermal exchange gas. The samples were measured in transmission using an area detector. The 2D data was integrated to 1D diffraction data utilizing the Fit2D software [25]. Corrections to obtain the $I(Q)$ and subsequent Fourier Transform with a $Q_{max}$ of 24 Å to obtain the $G(r)$ was performed using the program PDFgetX2 [26]. Analysis of the total scattering data was performed using the PDFgui software suite [27] over a range of 1.75 – 10.0 Å.

Bulk characterization of the magnetic and electronic properties was carried out for samples across the series. Measurements of the field and temperature dependence of the magnetic susceptibility were performed on powder samples (≤ 20 mg) mounted in polypropylene capsules. $\chi(T)$ and $M(H)$ measurements were collected using a MPMS3 Quantum Design SQUID magnetometer. Magnetization measurements at higher fields were taken using a vibrating sample magnetometer (VSM) mounted on a Quantum Design DynaCool Physical Properties Measurement System (PPMS) equipped with a 9 T magnet. Charge transport measurements were taken on well-sintered pellets using a four-probe configuration in a DynaCool PPMS. The samples were sanded into a bar geometry and mounted to the sample puck with GE varnish, while contacts were made using silver paste.

XAS and XMCD measurements at the Ir $L_{2,3}$ absorption edges were performed on beamline 4-ID-D at the APS. Powder samples (from the same batches used at 11-BM for the same $x$) were prepared by first sieving to a maximum particle size of 5 microns to prevent pin-holing. The sieved powder was then adhered to layers of tape which were stacked until a uniform sample thickness corresponding to roughly two absorption lengths was obtained. All measurements were performed using the transmission geometry. Energy selectivity was obtained using a double-crystal <111> Si monochromator while circularly polarized x-rays for the XMCD measurements were produced using a diamond crystal phase retarder optic in Bragg-transmission geometry, operated in helicity-switching mode at 13.1 Hz. In all cases, the absorption signal was detected using a diode with a lock-in amplifier [28]. In order to remove instrumental artefacts from the data, the XMCD measurements were performed with field ($\mu_0 H$ = 5 T) oriented both parallel and antiparallel to the incident wave-vector.

## 3. Results

### 3.1. Structural characterization

Data from synchrotron x-ray scattering experiments were collected on air sintered samples to determine the effect of increasing Ca-doping on the symmetry of the average and local structures. While neutron powder diffraction is a natural complement to this data, particularly for insight into oxygen-dependent behaviours, such an experiment was not carried out due to the high absorption cross section of both Ir and Eu [29]. The average pyrochlore structure is composed of interpenetrating networks of corner-sharing tetrahedra formed by the A (Eu) and B (Ir) site ions. In $Eu_2Ir_2O_7$, the AIAO magnetic order consists of Ir moments aligned along the local <111> axes of the tetrahedra and arranged such that, for each tetrahedra, all moments point either into or away from the center. The B-site ion is coordinated by a trigonally distorted octahedra of oxygen atoms, with the degree of distortion and Ir-O-Ir bond angle controlled by the $x$-coordinate of the $48f$ oxygen site ($u$). This is the only free structural variable besides the lattice parameter ($a$). An undistorted octahedron is attained for $u = .3125$. Fits to the $T = 300$ K synchrotron XRD data are shown in **figure 2**. At all measured temperatures, the patterns are well fit by the cubic $Fd\overline{3}m$ (#227) pyrochlore structure, with the additional presence of up to 2.5% Eu$_2$O$_3$ and $\leq$ 1.5% of both IrO$_2$ and Ir impurities. Comparison of lab XRD indicates that the vacuum-sintered samples tend to have a higher Eu$_2$Ir$_2$O$_7$ phase fraction. This is most likely driven by the second addition of IrO$_2$ prior to the vacuum sintering process, which compensates for volatilized IrO$_2$. The refined value of $u$ decreases marginally with Ca-doping, which results in a slight modification of the Ir-O-Ir bond angle. The refined value of $a$ also decreases with Ca-doping. A qualitatively similar trend for $u$ and $a$ was found from laboratory XRD on the vacuum-sintered samples.

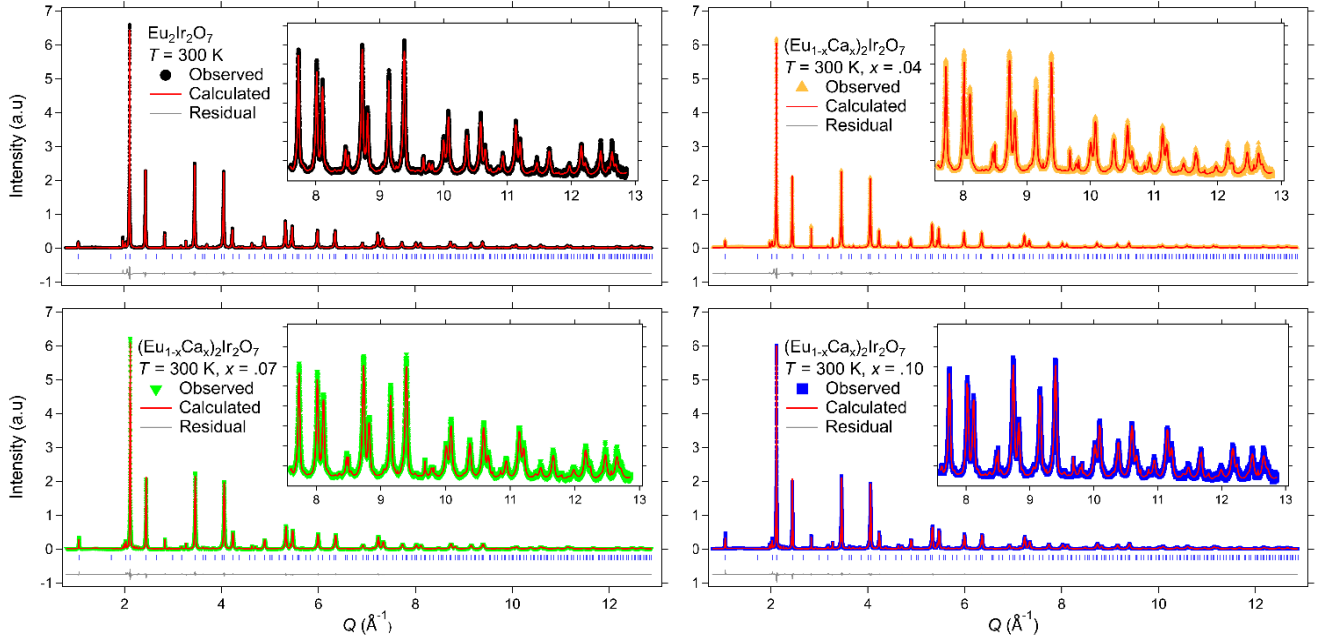

**Figure 2:** Rietveld refinements of synchrotron x-ray powder diffraction data for air sintered $x = 0$, .04, .07, and .10 samples taken at $T = 300$ K. The calculated curves include the pyrochlore phase and impurity phases (Ir, IrO$_2$, and Eu$_2$O$_3$). The blue ticks above the residuals indicate the location of pyrochlore peaks only. Other peaks come from the impurity phases (e.g. Ir at 2.8 Å$^{-1}$) as quantified and described in the text. Insets: High-$Q$ regions of the data and refinements.

Several different models were explored during the refinement process and the results of the final model are shown in table 1. A slight asymmetry toward low-$Q$ was observed for the Eu$_2$Ir$_2$O$_7$ peaks, as has been seen in other pyrochlore iridates [23] and which is most likely due to a distribution of strain in the lattice. This distribution was modelled by including a second Eu$_2$Ir$_2$O$_7$ phase with an increased lattice parameter (typically $\approx$ .1%) and a broad line shape, which significantly improved the fit. Comparison was made between fits with Ca on either the Eu or Ir site, and in all cases substitution of Ca for Ir produced inferior fits.

Due to the low scattering amplitude of the dilute Ca dopant, refinement of the Ca occupancy value proved unreliable. Subsequently, Ca occupancy was fixed to the $x$ value determined via WDXRF. For the parent sample, initial refinement of the Eu and Ir site occupancies showed complete occupancy. When site mixing was allowed between the Eu and Ir sites - such that any decrease in Eu (Ir) occupancy on the A (B) site was compensated by Ir (Eu), with total occupancy constrained to 1 - a maximum of 2% anti-site defects on the Ir sublattice was observed, with other values all < 1%. For the $x > 0$ samples, the Eu and Ir occupancies were initialized at the values of 1-$x$ and 1, respectively. The Eu site occupancy consistently refined to a value close to the nominal Eu stoichiometry (1 – $x_{nom}$ in **table 1**). All refinements of the $x > 0$ samples showed < .5% anti-site defects on both sublattices. Generally, this site-mixing is of the same magnitude as the reported errors and inclusion in the model only marginally improves the goodness-of-fit parameters. For the $x > 0$ samples, the refined Eu occupancy, when considered with the XRF-determined value of $x$, suggests that there is a variable concentration of vacancies on the A-site ranging from <1% ($x = .04$) to roughly 4% ($x = .10$)

The PDF data (**figure 3a, b**) show little change across the series and the data for the $x = 0$, .04, and .10 air-sintered samples taken at 300 K are all very similar. No new peaks or observable changes in peak shape in the G($r$) are observed with increased $x$ or between high and low temperatures i.e. as one cools below the magnetic transition or the MIT (if present). The data remain well fit to a model corresponding to the average pyrochlore structure found by XRD in this work and in the literature [30], out to the highest doping studied. The goodness of fit parameter, $R_w$, falls between 9% and 15% for all fits. Results from these refinements are shown in **figure 4**. The A$_2$B$_2$O$_7$ structure allows only one pairwise metal-to-metal (M-M) length even as the M-O-M angles change. As such, the refined M-M distance is an average over all pairwise combinations of M-M nearest neighbors (Eu-Eu, Ir-Ir, Ca-Ir etc.). The dependence of the M-M distance on temperature is roughly linear for each sample, showing a small decrease in parallel with the lattice parameter, consistent with the average structure data. No strong feature is observed at $T_{MIT}$ or $T_N$ although the $x = 0$ and .04 samples do show a slight deviation (within error) from the linear decrease.

**Table 1:** Results from Rietveld refinement of the synchrotron x-ray powder diffraction data. $X_{nom}$ and $X_{XRF}$ are the nominal and quantitative WDXRF values of the Ca-substitution, respectively. All $x$ values in the text refer to the the $x_{XRF}$ value. The data (by row) are as follows: $a$ = pyrochlore cubic lattice parameter, $u$ = $x$-position of the O 48$f$ site, angles within the IrO$_6$ octahedra, Eu and Ir occupancy values on the A- and B-sites (A$_2$B$_2$O$_7$), $U_{iso}$ = isotropic thermal parameters for the A- and B-sites, pyrochlore phase fraction (weight %), Rietveld goodness of fit parameters. At all times during the refinement, oxygen occupancies and $U_{iso}$ values were fixed to 1 and .001, respectively. Further details on the refinement model are provided in the text.

| $x_{nom}$ | 0 | 0.05 | 0.1 | 0.15 | 0 | 0.05 | 0.1 |
|---|---|---|---|---|---|---|---|
| $x_{XRF}$ | 0 | 0.044(1) | 0.071(1) | 0.102(1) | 0 | 0.044(1) | 0.071(1) |
| T | **300 K** | | | | **90 K** | | |
| $a$ (Å) | 10.28330(3) | 10.28076(3) | 10.27588(3) | 10.27004(2) | 10.27381(2) | 10.27107(3) | 10.26563(2) |
| $u$ (O$_{48f}$) | .3341(2) | .3334(2) | .3321(1) | .3316(1) | .3354(2) | .3343(2) | .3340(2) |
| ∠ Ir-O-Ir (°) | 129.1(2) | 129.5(2) | 130.2(2) | 130.4(3) | 128.4(4) | 129.0(3) | 129.2(3) |
| ∠ O-Ir-O (°) | 81.9(5) | 82.1(3) | 82.6(4) | 82.8(1) | 81.5(1) | 81.8(3) | 81.9(3) |
| Eu, A occ. | .996(1) | .948(1) | .903(4) | .857(4) | .996(2) | .950(1) | .901(1) |
| Ir, A occ. | .004(1) | .002(3) | .001(4) | .000(3) | .004(2) | .001(1) | .003(1) |
| Ir, B occ. | .992(1) | 1.000(4) | 1.000(3) | .995(3) | .979(2) | 1.000(1) | 1.000(1) |
| Eu, B occ. | .008(1) | .000(4) | .000(3) | .005(1) | .021(2) | .000(1) | .002(1) |
| A, $U_{iso}$ (Å$^2$) | .00736(6) | .00747(3) | .00734(3) | .00826(3) | .00405(3) | .00333(3) | .00424(8) |
| B, $U_{iso}$ (Å$^2$) | .00419 (4) | .00384(1) | .00260(4) | .00183(1) | .00255(1) | .00114(1) | .00108(5) |
| Eu$_2$Ir$_2$O$_7$ (%) | 94.71 | 95.52 | 94.8 | 94.49 | 94.44 | 95.6 | 97.39 |
| $R_{wp}$ (%) | 8.69 | 8.35 | 7.45 | 7.31 | 8.84 | 8.50 | 8.28 |
| $\chi^2$ | 1.94 | 1.89 | 1.67 | 1.66 | 1.99 | 1.88 | 1.53 |

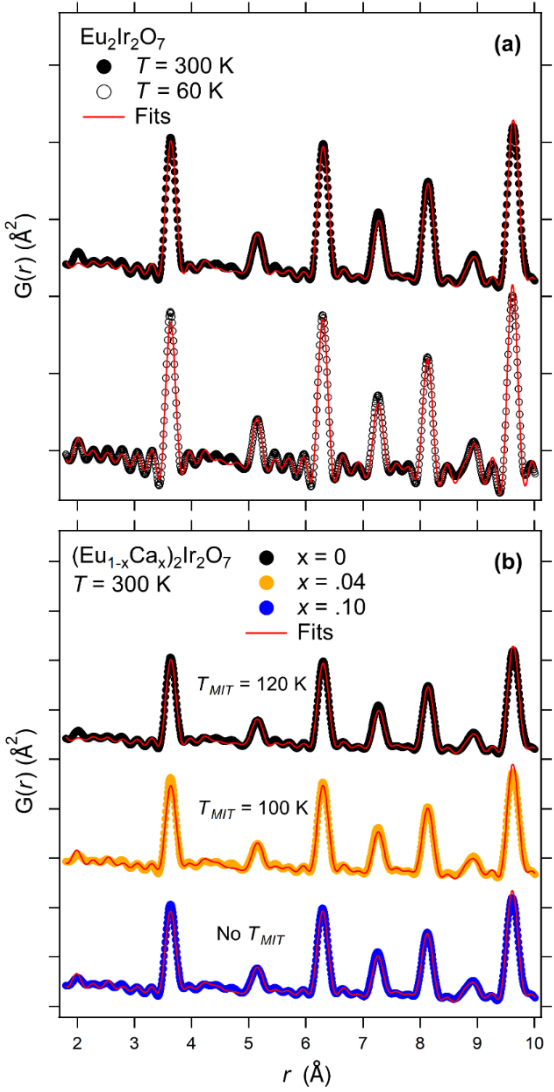

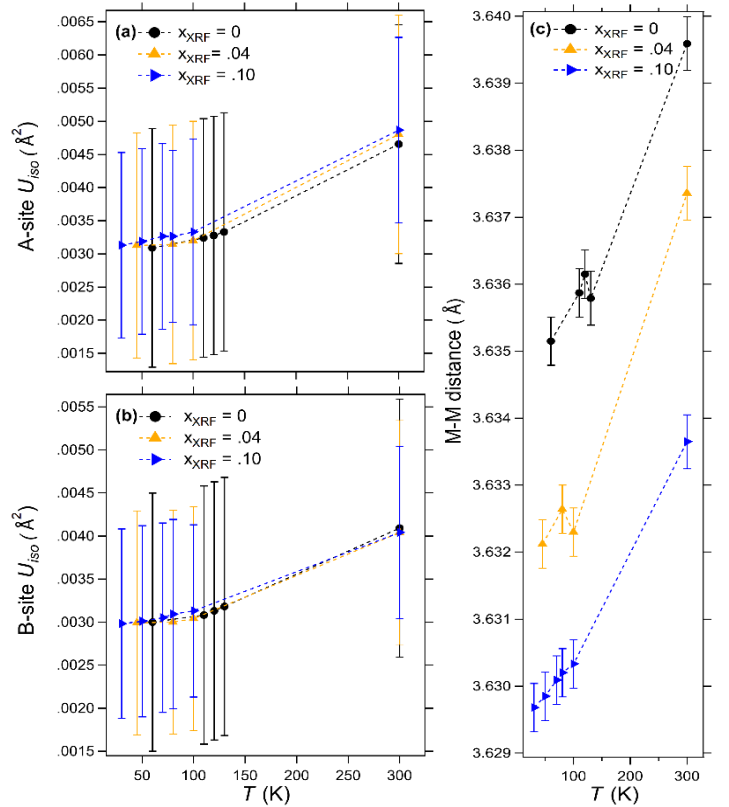

**Figure 3**: PDF refinement of synchrotron x-ray total scattering data on air-sintered samples. Curves offset for clarity (**a**) Refinements for $x = 0$ at $T = 300$ (top, $R_W = 9.6\%$) and 60 K (bottom, $R_W = 14.1\%$) with $T_{MIT} = 120$ K (**b**) Refinements for $x = 0$ ($R_W = 9.6\%$), .04 ($R_W = 10.6\%$) and .10 ($R_W = 9.3\%$) at $T = 300$ K.

The isotropic atomic displacement parameters for Ir and Eu remain small and of similar magnitude for all $x$ values studied and decrease with temperature as expected.

### 3.2 Charge transport and magnetism

We now turn to the changes observed in the charge transport and magnetism that accompany the introduction of Ca onto the Eu-site. The resistivity measurements on vacuum-sintered samples (**figure 5** – results for air sintered samples can be found in the **Appendix**) show that $T_{MIT}$, identified via the inflection in $d\rho/dT$, is fully supressed within the doping range $.04 < x < .07$. This is in agreement with a previous report on $(Eu_{1-x}Ca_x)_2Ir_2O_7$ [19]. At $x = .04$ the upturn in $\rho(T)$ is considerably less sharp and weakened compared to the parent sample, and the insulating state is less well-defined. This is quantified by the inverse relative resistivity ratio, $1/RRR = \rho(2K) / \rho(300K)$, which decreases significantly from $1/RRR \approx 2500$ for the parent, to $1/RRR \approx 10$ for $x = .04$. The resistivity takes a clear metallic form for $x = .07$ and $.13$, with no upturn observed down to 2 K. In the nominally metallic regime ($T > T_{MIT}$) of parent $Eu_2Ir_2O_7$, both air and vacuum-sintered samples show $d\rho/dT < 0$ (incoherent) transport behavior as has been reported previously for both polycrystalline and single crystal samples [2, 16]. Ca-doping leads quickly to $d\rho/dT > 0$ at high temperatures, reminiscent of the crossover behaviour observed under increasing hydrostatic pressure [16], as is illustrated by $\rho(T)$ for the $x = .02$ air sintered sample (see **Appendix**).

**Figure 4:** Results from PDF refinement of synchrotron x-ray total scattering data on air-sintered samples at various temperatures (**a**) A-site atomic displacement parameters (**b**) B-site atomic displacement parameters (**c**) Metal-to-metal (M-M) bond distance.

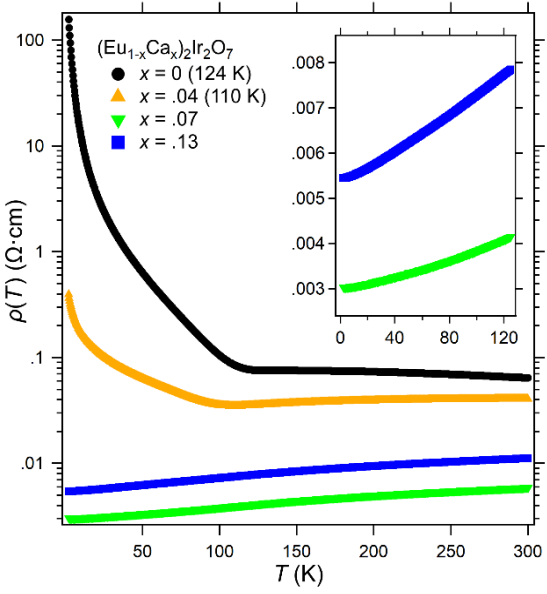

**Figure 5:** Resistivity measurements on pellets of vacuum-sintered samples. All data were taken on warming from 2 K to 300 K. The values in parenthesis are $T_{MIT}$, determined as described in the text. Inset shows the low temperature ($T < 125$ K) behavior of the metallic samples.

Magnetic susceptibility measurements (**figure 6**) show the presence of a weak irreversibility in all samples. This irreversibility is associated with the emergence of the AIAO order on the Ir sublattice, as demonstrated by muon spin relaxation/rotation [31] and resonant x-ray diffraction [3] experiments on the parent material. Although the mechanism for the irreversibility has not been unambiguously determined, it is typically attributed to a slight canting of the tetrahedral arrangements of Ir spins [32, 33]. $T_{AFM}$ is therefore best visualized by the subtraction $\chi_{FC} - \chi_{ZFC}$ shown in **figure 6b** and is defined here by the onset of irreversibility (i.e. the point where the $\chi_{FC} - \chi_{ZFC}$ curve increases rapidly). While the MIT and AIAO ordering occur together in the parent sample ($T_{AFM} = T_{MIT} = 124$ K), the transitions quickly become decoupled in the doped samples. $T_{AFM}$ shows only a small decrease even as $T_{MIT}$ is fully suppressed. $M(H)$ data for both air and vacuum sintered samples (not shown) give a linear response at all temperatures and all doping levels for fields up to $H = 9$ T, with no clear indication of saturation or significant hysteresis. The magnetism is largely dominated by the Van Vleck susceptibility of the Eu$^{3+}$ ion, which is roughly 75% of the total signal. Subtraction of the Van Vleck term (discussed later in this section) did not qualitatively change the form of the *M(H)* curves.

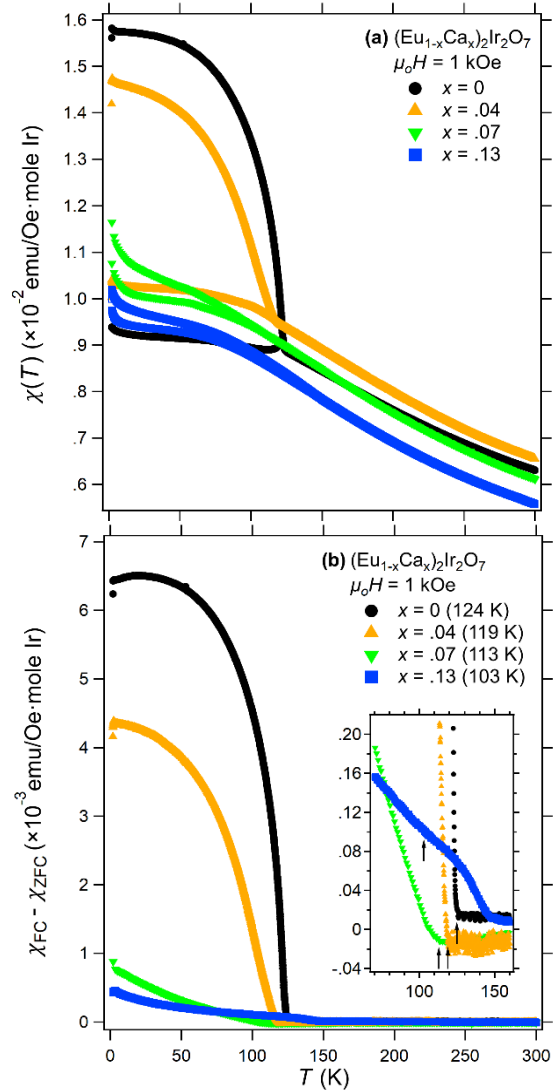

**Figure 6:** (a) Magnetic susceptibility, $\chi = M/H$ versus temperature data for vacuum-sintered samples. All curves are taken on warming from 2 K to 300 K with a field of 1 kOe. The lower curve is taken after cooling in zero field (ZFC) and the upper curve is taken after cooling in a field of 1 kOe (b) Subtraction of the ZFC curve from the FC one, used to represent the evolution of the irreversibility with Ca-substitution. The values in parenthesis are $T_{AFM}$ as described in the text. **Inset:** Closer view of $\chi_{FC} - \chi_{ZFC}$ in the region around $T_{AFM}$ (arrows), as defined in the main text.

While increased Ca-doping decreases the difference between the $\chi_{FC}$ and $\chi_{ZFC}$ curves, this irreversibility surprisingly persists into the regime where $\rho(T)$ is metallic, as seen clearly in the *x* = .07 sample. Direct determination of $T_{AFM}$ is difficult for the metallic *x* = .13 sample because the irreversibility data is obscured by a slight, extrinsic upturn in $\chi_{FC} - \chi_{ZFC}$ at 150 K. This 150 K feature is seen much more clearly in the air sintered samples, even at significantly lower doping levels, and dominates the overall susceptibility at higher doping. This is a small moment, synthesis-dependent impurity effect that is dramatically reduced in vacuum sintered samples. Nevertheless, this impurity effect remains weakly resolvable in the *x* = .13 sample where the primary magnetic order is almost quenched. As a result, $T_{AFM}$ for the *x* = .13 sample cannot be extracted by direct examination of the $\chi_{FC} - \chi_{ZFC}$ curve and has been estimated by linearly extrapolating from the lower doping values. Examination of the first derivative of the $\chi_{FC} - \chi_{ZFC}$ curve supports this extrapolated value, though it remains an estimate. We note here that, other than this 150 K transition, no qualitative difference is observed between the air and vacuum sintered samples (i.e. the charge transport and structural properties). Further details on this point are provided in the **Appendix**, where we argue that the

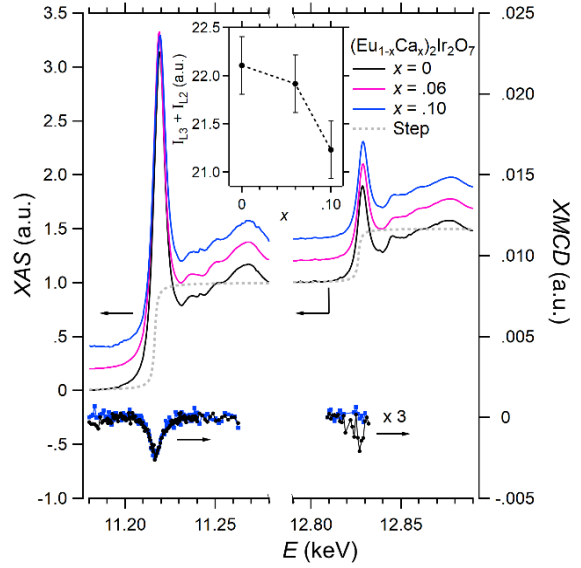

**Figure 7:** X-ray spectroscopy measurements collected on air-sintered samples. The solid lines correspond to the XAS data (*T* = 300 K) while the markers show the XMCD (*T* = 5 K, 1 T field-cool, 5 T measurement). The dashed line is the step function used in the XAS analysis. Inset: Sum of the $L_2$ and $L_3$ white-line integrated

extra 150 K feature in the susceptibility is due to a secondary state within the sample.

### 3.3 XAS and XMCD

X-ray absorption spectroscopy (XAS) and magnetic circular dichroism (XMCD) provide an element-specific means of probing the spin-orbit coupled nature of the $5d^5$ states of Ir$^{4+}$. XAS and XMCD spectra collected at the Ir $L_2$ ($2p_{1/2} \rightarrow 5d_{3/2}$) and $L_3$ ($2p_{3/2} \rightarrow 5d_{3/2,5/2}$) absorption edges for a range of air sintered samples are shown in **figure 7**. Select results from the analysis are presented in **table 2**. In the XAS data, a clear enhancement of the integrated intensity of the white-line is seen for the $L_3$ edge relative to the $L_2$ edge ($I_{L2,L3}$, a description of the calculation can be found in Porter *et al.* [18], following Clancy *et al.* [34]). This increase is quantified by the branching ratio, $BR = I_{L3}/I_{L2}$, which shows a significant increase over the "statistical" value of $BR = 2$, though it decreases with Ca-doping. The statistical value is expected for states with negligible SOC, where the ratio of $I_{L3}$ and $I_{L2}$ is dependent only on the relative numbers of initial core electron states [35, 36, 37]. Due to dipole selection rules in the XAS process ($\Delta J = 0, \pm 1$), the enhancement of the *BR* indicates a predominant $J = 5/2$ character for the 5*d* hole states. Thus, deviation from the statistical *BR* indicates a significant influence of SOC, which reconstructs the $t_{2g}$ manifold [1, 34, 38]. These observations are in-line with expectations for the $J_{eff} = ½$ state [1, 34]. The influence of SOC can be further seen in the expectation value of the spin-orbit operator, $<L·S>$, which can be directly related to the *BR* via the relation $BR = (2+r)/(1-r)$ and $r = <L·S> / n_h$ where $n_h$ is the average number of 5*d* holes (equal to 5 for $Ir^{4+}$) [35, 36, 37, 38]. While not equivalent to the SOC term in the Hamiltonian (**H**$_{SOC}$ = ζ**L**·**S**), the large value of $<L·S>$ observed here indicates the persistent influence of substantial SOC and that the orbital component of the angular momentum has not been quenched [34]. The XMCD measurement at 5 K (**figure 7**) shows a small, but well-resolved signal, at the $L_3$ edge, with roughly .3% difference in absorption between the two x-ray helicities. A very weak signal is observed at the $L_2$ edge for the parent sample, while there is no signal resolvable for the $x = .10$ sample. The occurrence of large Ir $L_3$ XMCD signal with minimal $L_2$ signal in both samples is characteristic of the $J_{eff} = ½$ state. The persistence of this behavior in the metallic $x = .10$ sample contrasts with the nearly zero Ir $L_2$ and $L_3$ XMCD signals seen for metallic/paramagnetic $Pr_2Ir_2O_7$ [39]. This disparity highlights that carrier-doping provides a distinct mechanism from chemical pressure for modifying $T_{AFM} / T_{MIT}$ and indicates a tangible distinction between the resulting metallic states.

**Table 2:** Results from analysis of the XAS (T = 300 K and *x* = 0, .06, .10) and XMCD (T = 5 K and *x* = 0, .10) at the Ir $L_{2,3}$ edges. The nominal $n_h$/valence is determined by the *x* value. The XAS $n_h$/valence is calculated by multiplying the valence in the parent (4+) by the ratio of $I_{doped}/I_{parent}$, where $I = I_{L2} + I_{L3}$. BR = branching ratio, <*L·S*> = expectation value of the spin-orbit operator, $m_{tot}$ = sum of the spin ($m_s$) and orbital ($m_l$) moments at H = 5 T, $L_z/S_z = m_l/m_s$

| *x* | $n_h$ ($q_e$) | | Ir valence | | BR (a.u.) | <*L·S*> (ℏ$^2$) | $m_{tot}$ (μB/Ir) | $L_z/S_z$ |
|---|---|---|---|---|---|---|---|---|
| | Nominal | XAS | Nominal | XAS | | | | |
| 0 | 5 | — | 4+ | — | 5.7(3) | 2.77(9) | .0045(9) | 3.3 |
| 0.06 | 5.06 | 4.97(8) | 4.06+ | 3.97(8)+ | 5.6(2) | 2.70(9) | — | — |
| 0.1 | 5.08 | 4.84(8) | 4.10+ | 3.84(8)+ | 4.8(2) | 2.32(9) | .0053(6) | 2.3 |

Calculations based on application of the sum rules reveal a small Ir moment of $\approx .005$ $\mu_B$/Ir. Details on this calculation and estimation of the $<T_Z>$ term can be found in Porter *et al.* [18]. A previous measurement for $Sr_2IrO_4$, which also contains $Ir^{4+}$ in a cubic CEF environment, showed a $L_3$ XMCD signal of 3%, with a corresponding net moment of $\approx .05$ $\mu_B$/Ir in the weakly ferromagnetic state [40]. This scales well with our own results. The calculated ratio $L_Z/S_Z = 3.3$ and 2.4 for $x = 0$ and .10, respectively, indicates a large orbital contribution to the total moment and the positive sign indicates parallel spin and orbital moments (unlike what is expected for the atomic $J = ½$ state [1]). The decrease with doping is driven by the change in the $L_2$ XMCD and is likely exaggerated since an $L_2$ signal was not well resolved in the measurement of the $x = .10$ sample (note that the $L_z/S_z$ ratio is independent of the number of holes). These results, together with the enhanced *BR* from the XAS data, provide a clear indication of a $J_{eff} = ½$ state in $Eu_2Ir_2O_7$, which is gradually weakened with Ca-doping and the onset of metallic behavior but is not completely destabilized.

The calculated moment from XMCD ($\approx .005$ $\mu_B$/Ir) is similar to values found from bulk DC magnetometry measurements (magnetization versus field) at $T = 5$ K for other pyrochlore iridates with non-magnetic A-sites: $Y_2Ir_2O_7$ (.005 $\mu_B$/Ir) [21] and $Lu_2Ir_2O_7$ (.01 $\mu_B$/Ir) [41] at $H = 5$ T. Reported magnetization measurements for $Eu_2Ir_2O_7$, as well as our own, show a net magnetization of .09 $\mu_B$/Ir at 5 T [20, 42]. The origin of the discrepancy is due to the Van Vleck contribution of $Eu^{3+}$, which does not contribute to the Ir XMCD. To deconvolve the Van Vleck and AIAO contributions, we measured χ(T) for a similarly prepared sample of $Eu_2Ir_2O_7$ under the same field conditions as that used for the XMCD experiment (1 T field-cool, 5 T measurement field). The data above $T_{AFM}$ (125 K < T < 300 K) were fit to the form χ$_{VV}$(T) + χ$_0$ where χ$_{VV}$ is the Van Vleck contribution with a single fitting parameter (λ, the spin-orbit coupling constant) [43]. $\chi_0$ is a (positive) constant Pauli paramagnetic term and was constrained to be close to the essentially temperature independent susceptibility above $T_{AFM}$ seen for $Y_2Ir_2O_7$ and $Lu_2Ir_2O_7$ (≈ 5 x $10^{-4}$ emu/mole Ir·Oe ) [21, 41]. After subtracting the Van Vleck contribution, the moment from DC magnetometry at $T$ = 5 K is ≈ .008 $\mu_B$/Ir, in reasonable agreement with the XMCD moment.

Due to the sensitivity of the XAS process to unoccupied 5$d$ states, carrier-doping can be quantified by analysis of the white-line features [34, 38]. Since the sum $I_{L2} + I_{L3}$ is proportional to the local density of unoccupied 5$d$ states, comparison of Ca-doped samples to the parent compound reflect changes in the average Ir valence. Generally, this change can also be seen through shifts of the energy of the white-line, with a positive shift indicating an increase in valence. Here we see only small shifts in the white-line position for our Ca-substituted samples (on the order of .1 eV), which do not show a consistent direction. The small shift at these doping values is in line with the observed change of 1.3 eV/hole from a survey of 4+, 5+ and 6+ iridates [34]. Considering the energy resolution and repeatability of the monochromator (1 and .1 eV, respectively) the lack of a consistent trend in the white line shifts is not surprising. Since the substitution of $Ca^{2+}$ for $Eu^{3+}$ should result in the addition of holes to the originally half-filled $J_{eff}$ = ½ state of $Ir^{4+}$ the value of $I_{L2} + I_{L3}$ is also expected to increase. Note that in the strong SOC limit, holes in the $J_{eff}$ = ½ state are accessed preferentially by the $L_3$ edge since those states are derived from the $J$ = 5/2 states of the SOC split 5$d$ manifold [1]. As such $I_{L3}$ is much more sensitive to changes in the $J_{eff}$ = ½ hole population than $I_{L2}$ and they need not change in tandem. However, we observe a marked *decrease* in $I_{L3}$ from the parent to the $x$ = .10 sample. The observed trend in $BR$ is also opposite what would be expected for hole-doping, though the magnitude of the change is greater than would be expected from the small changes in Ir valence.

These results are suggestive of electron doping rather than the expected hole doping.

## 4. Discussion

Clear changes in the structure and magnetism accompany the breakdown of the SOC-assisted Mott insulating state across the series $(Eu_{1-x}Ca_x)_2Ir_2O_7$. The structural evolution with increasing Ca-doping indicates that the crystallographic symmetry of the parent compound is maintained across the series and below $T_{MIT}$ for both the average (long-range) and local (short-range) structures. Although the symmetry of the average structure is maintained with Ca-substitution, features of the structure are still modified. First, $u$ decreases with increasing Ca-content, leading to a slight increase of the Ir-O-Ir bond angle. This observation implies that that Ca-doping leads to a small, sterically driven increase in the Ir valence bandwidth due to greater overlap of the Ir 5$d$ $t_{2g}$ orbitals [44]. However, the value of the Ir-O-Ir bond angle remains less than the value found for metallic $Pr_2Ir_2O_7$ (≈132°) [45] for all samples.

Counter-intuitively, despite 8-fold coordinated $Ca^{2+}$ being larger than $Eu^{3+}$, the other structural change induced by Ca-doping is a decrease in the lattice parameter. This behaviour is consistent with results from a previous study [19] and the fact that an early report of $Ca_2Ir_2O_7$ shows a lattice parameter significantly smaller than $Eu_2Ir_2O_7$ [46]. Even though the bond angle change is analogous to what is seen upon moving to a larger A-site cation, the lattice parameter changes in the opposite direction. This indicates that the structural response to Ca-doping is not simply dependent on the difference in ionic radius relative to the ion that is being replaced (i.e. a steric effect), but other effects such as an electronic deformation potential may dominate the lattice response. The resulting reduction in interatomic distances may also contribute to a small increase in the Ir valence bandwidth. However, the weak dependence of $T_{MIT}$ on pressure [16], which effects both the Ir-O-Ir bond angle and lattice parameter [47], means that these small changes in bandwidth most likely do not contribute meaningfully to the depression of $T_{MIT}$. Rather, filling-control is seen as the dominant cause of metallic behavior in (Eu$_{1-x}$Ca$_x$)$_2$Ir$_2$O$_7$, as discussed later. The small increase in the bandwidth may still be partially responsible for the accelerated decrease of the *BR*, since it acts to mix the $J_{eff}$ states [48].

The XAS data indicate a clear change in the Ir valence accompanies the depression of $T_{MIT}$ with Ca-doping. However, the nature of this filling-control is, based on the change in the white-line intensities, a net increase in the number of electrons in the Ir $J_{eff} = \frac{1}{2}$ state (i.e. electron doping). Based on the robust Ca$^{2+}$ oxidation state this suggests an active compensatory mechanism for the holes added by Ca-doping. A bond valence sum (BVS) calculation for Ca$^{2+}$ in (Eu$_{1-x}$Ca$_x$)$_2$Ir$_2$O$_7$, using the Ca/Eu-O distances derived from the results of the synchrotron XRD refinement, shows that Ca is consistently overbonded (i.e. the calculated BVS value for 8-fold coordination is greater than the formal valence of 2+). Such overbonding can be a sign of compensating defects or displacements within the oxygen sublattice [49]. Oxygen vacancies, which are known to occur readily in the pyrochlore structure [44, 50, 51] are therefore a natural candidate for the proposed compensatory mechanism. The presence of A-site vacancies – suggested by the sum of the XRF determined value of the A-site Ca occupancy and the refined value of the A-site occupancy from synchrotron XRD being < 1- should also favor compensating oxygen vacancies.

Unfortunately, characterizing the O-site occupancy in Eu$_2$Ir$_2$O$_{7-y}$ is challenging as x-rays lack relative sensitivity to oxygen and, as previously mentioned, absorption issues impede a high-resolution application of powder neutron diffraction. A simple calculation based on charge neutrality with the A-site vacancy concentration indicated by the synchrotron XRD refinement and the XAS determined Ir valences yields *y*-values of .18 and .56 for the *x* = .06 and .10 samples, respectively. This rough estimate of the oxygen content is on par with what is found for (Bi$_{1-x}$Ca$_x$)$_2$Ir$_2$O$_{7-y}$ synthesized in air, where powder neutron diffraction shows *y*-values of ≈ .1 - .2 [49, 51], although no comparable relationship between *x* and *y* was noted previously [49]. This indicates that, if *y* scales with *x*, electron-doping via oxygen vacancies can dominate over the effect of Ca-doping.

It should be noted that the parent iridates do not seem to suffer a propensity to form oxygen vacancies. Therefore, other possible explanations for the XAS data should also be considered. Since the Ir-O bond lengths in the IrO$_6$ octahedra depend directly on the value of the lattice parameter and *u*, Ca-doping also shortens these bonds. Therefore, an alternative explanation for the XAS data is that these shortened bond lengths enhance the hybridization of the O 2*p* and Ir 5*d* orbitals (i.e. increase covalency), causing compensatory charge transfer to the Ir site. BVS analysis provides support for increased covalency of the Ir-O bond. In Eu$_2$Ir$_2$O$_7$, the Ir$^{4+}$ BVS is found to be 4.07, relatively close to the formal valence. For Ir$^{5+}$ in Ca$_2$Ir$_2$O$_7$ [46] the value (6.33) is substantially higher that the formal valence, consistent with increased covalency. Charge transfer has also been invoked to explain the decrease in the XAS $I_{tot} = I_{L2} + I_{L3}$ with increasing (physical) pressure observed for the hyper-kagome iridate compound Na$_3$Ir$_3$O$_8$ [48]. In the case of Na$_3$Ir$_3$O$_8$, the change in Ir-O bond distance is relatively small (≈ .1% from ambient to 10 GPa) yet a significant (≈ 10%) decrease in $I_{tot}$ is observed. In Eu$_2$Ir$_2$O$_7$, the change in Ir-O bond length is significantly larger (changing roughly linearly with ≈ .65% difference between *x* = 0 and .10) while the change in $I_{tot}$ is less (≈ 4%). The difference in magnitude could potentially be accounted for by the fact that the added holes from Ca-doping are driving $I_{tot}$ in the opposite direction. While we cannot directly determine whether one, or both, of these compensatory mechanisms causes the observed valence change, these results indicate that direct determination of the Ir valence should be considered an essential component in interpreting doping studies of pyrochlore iridates.

Returning to the x-ray data, the total scattering measurements provide a means for examining whether changes in the local structure are relevant to the mechanism of the MIT. Changes in local structure can dramatically modify electronic and magnetic behaviours, as has been studied in the manganite perovskites [54, 55, 56]. However, the consistency of the PDF spectra for $Eu_2Ir_2O_7$ across the thermal $T_{MIT}$ (**figure 3a**) does not support this idea. The lack of new correlation peaks, changes in peak shape at low temperature, or significant modification of the high temperature peaks, indicates that the symmetry of the local structure at the A and B-sites does not deviate at $T < T_{MIT}$ (within resolution) from that of the average structure. The similarity of the room temperature spectra for all samples (**figure 3b**) further indicates that the addition of Ca does not alter the local structure symmetry.

All spectra of doped compounds remain well fit by the crystallographic structure of the parent, and this similarity suggests that Ca enters the lattice homogeneously without observable nanophase chemical separation. If inhomogeneous local distortions were occurring with doping, then one would expect broadening of peaks associated with the M-M distances or formation of a new correlation peak due to the different nearest neighbor distance for a Ca-Ca pair. This is not observed. Focusing on the highest doped sample ($x = .10$), where local distortions would be most readily detected, we see that the data remains well described by a disordered, high-symmetry model. The M-M distances (**figure 4c**) show only a small, roughly linear, decrease with doping with no distinct feature at $T_{MIT}$. The values and trend agree well with the changes in the average structure refined from the synchrotron XRD data. The lack of dependence of the isotropic atomic displacement parameters on doping, essentially unchanged within error, further supports the picture of homogeneous carrier-doping and suggests minimal disorder of the local structure. However, the data presented here is not sensitive enough, at this level of doping, to rule out changes in the local structure around the Ca ions. It remains possible that oxygen and/or A-site vacancies occur preferentially near the Ca or that the overbonding of the Ca results in neighboring displacements on the oxygen sublattice.

The approach to a fully metallic state is characterized by two related features in the magnetization: the decoupling of $T_{AFM}$ from $T_{MIT}$ and the persistence of the irreversibility in $\chi(T)$. This irreversibility coincides with the formation of the AIAO state confirmed for the parent compound [3] and thus provides an easily observed metric for the evolution of the magnetism. The decrease in the magnitude of the irreversibility is therefore interpreted as an indication of the gradual breakdown of the AIAO state, with likely short-range AIAO correlations remaining at intermediate (metallic) levels of Ca-doping. The persistence of this signature in the metallic state suggests the occurrence of electronic phase separation, without, as evidenced by the total scattering experiments, observable corresponding chemical phase separation. Initial muon spin relaxation ($\mu^+$SR) experiments on $(Eu_{1-x}Ca_x)_2Ir_2O_7$ also offer some support for this picture, as the observed magnetic volume fraction decreases with increased Ca-doping [57]. Additionally, a separate $\mu^+$SR study on isovalent substitution in $(Eu_{1-x}Bi_x)_2Ir_2O_7$ reported an eventual cross-over to short-range magnetic order [58]. Analogous behavior has also been observed for electron-doped $(Sr_{1-x}La_x)_2IrO_4$, where the separation between $\chi_{FC}$ and $\chi_{ZFC}$ below the AFM transition is similarly depressed with doping and persists even as the charge gap is destabilized [59]. In this present case, the addition of charge carriers in $(Eu_{1-x}Ca_x)_2Ir_2O_7$ results in local regions where the $J_{eff} = ½$ state deviates from half-filling and the Mott gap collapses locally. This behavior is summarized in the phase diagram shown in **figure 8.**

As noted earlier, different relative evolutions of $T_{AFM}$ and $T_{MIT}$ have been reported in similar doping studies. The observation for $(Y_{1-x}Ca_x)_2Ir_2O_7$ of $T_{AFM}$ showing no decrease with doping [21] is most likely due to the obscuring, extrinsic effect of the synthesis dependent high-temperature transition. Interestingly, $(Eu_{1-x}Ca_x)_2Ir_2O_7$ powder samples produced using high-pressure synthesis [19] do not show the decoupling of $T_{MIT}$ and $T_{AFM}$ observed in our study. Instead, $T_{MIT}$ and $T_{AFM}$ decrease in tandem. The origin of this discrepancy merits further investigation and comparison of the chemical and crystallographic details between powders synthesized under ambient and high-pressure conditions; however, at a minimum, it reveals that the coupling between the two transitions is sensitive to subtle structural details. Notably, for $(Nd_{1-x}Ca_x)_2Ir_2O_7$, synthesized in the same manner as reported here, $T_{MIT}$ and $T_{AFM}$ remained locked together. Therefore, it appears that the magnetism (or lack thereof) of the A-site ion may determine whether the compound can enter a similar phase separated state. Determining the root cause of this variability – as related to both synthesis method and the magnetism of the A-site ion - seems likely to shed light on the connection between the MIT and the AIAO order in this materials class.

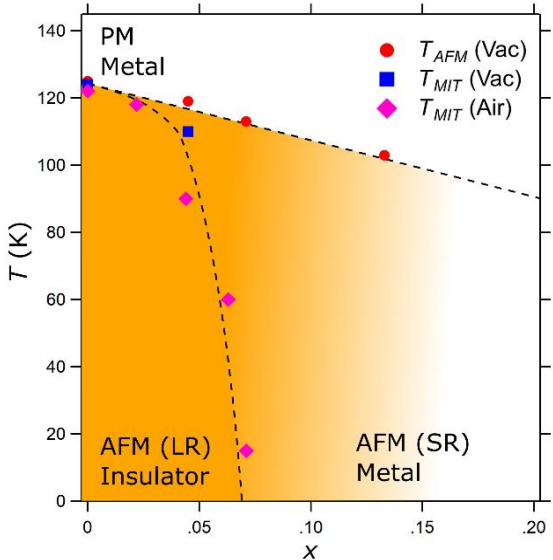

**Figure 8:** Phase diagram of $(Eu_{1-x}Ca_x)_2Ir_2O_7$ based on bulk magnetization ($T_{AFM}$) and charge transport measurements ($T_{MIT}$). $T_{MIT}$ is shown for both vacuum and air-sintered samples. Dashed lines indicate suggested phase boundaries while the color gradient represents the gradual evolution from long-range (LR) to short-range (SR) antiferromagnetism, followed by the eventual collapse of the AIAO state.

Finally, we consider again the small moment determined from the Ir $L$ edge XMCD measurement. It should be emphasized that, while XMCD probes the net ferromagnetic moment of the system, there are still two distinct mechanisms which may contribute to the measured signal: net ferromagnetism concomitant with the formation of AIAO order and/or a field-dependent response of the AIAO order to the field. More specifically, the former contribution could be weak ferromagnetism arising from the AFM domain walls (present due to the existence of "all-out-all-in" domains [60, 61]) while the latter could be spin-canting within the AIAO structure. That the measured Ir moment is consistent with other pyrochlore iridates with non-magnetic A-sites (A = Y, Lu) points to a common basis for the Ir XMCD signal. We also note that a parallel XMCD measurement on $Nd_2Ir_2O_7$ yielded a very similar value [18]. The source of this signal remains an open question, and XMCD measurements of the remanence (i.e. hysteresis loops) will be needed to distinguish between these two contributions. However, it is notable that the moment is identical within error upon progressing from an insulating ($x = 0$) to a metallic ($x = .10$) state. Regardless of the mechanism, this stability seems consistent with a picture of electronic phase separation: local pockets of the parent insulating/AIAO state, in which the weak Ir magnetic polarization is unchanged, remain deep into the metallic regime.

## 5. Conclusions

Our data indicate that substitution of Ca for Eu in $(Eu_{1-x}Ca_x)_2Ir_2O_7$ occurs without an observable change in lattice or site symmetry. $T_{MIT}$ is depressed with Ca-doping, and changes in the Ir valence and the Ir-O-Ir bond angle/Ir-O distance show that the low-temperature insulating state collapses under the majority influence of filling-control. The overall nature of the filling-control on the Ir-sites, contrary to expectation, is electron doping, suggesting that a significant compensatory mechanism is at work. The presence of oxygen vacancies, which scale with Ca-doping, or, alternatively, increased covalency of the Ir-O bonds are suggested as possibilities. The AIAO ordered state survives locally well into the metallic state, with $T_{AFM}$ responding only weakly to Ca-doping. This behavior is understood to be the result of the formation of an electronically phase separated state with carrier-doping. The dependence of this phase separation on the synthesis method, as well as other variability in bulk properties of doped and un-doped $Eu_2Ir_2O_7$, illustrates that even minor stoichiometry variations can dramatically alter the observed properties. Careful attention to synthesis and bulk characterization will be needed to ascertain the intrinsic properties of the pyrochlore iridates and realize unconventional states within them. Lastly, XMCD measurements reveal a weak Ir signal that appears to be a generic property of the Ir sublattice, although further measurements are needed to determine its origin.

**Appendix**

We provide here a brief discussion of the magnetism and charge transport of the air-sintered samples and their dependence on synthesis method. The main qualitative difference between the air and vacuum sintered samples is the occurrence of additional irreversibility at $T' = 150$ K in $\chi(T)$, above $T_{AFM}$. **Figure A** shows clearly that this is a synthesis dependent feature which is reduced upon vacuum sintering. It should be noted that this feature is also not seen in $(Eu_{1-x}Ca_x)_2Ir_2O_7$

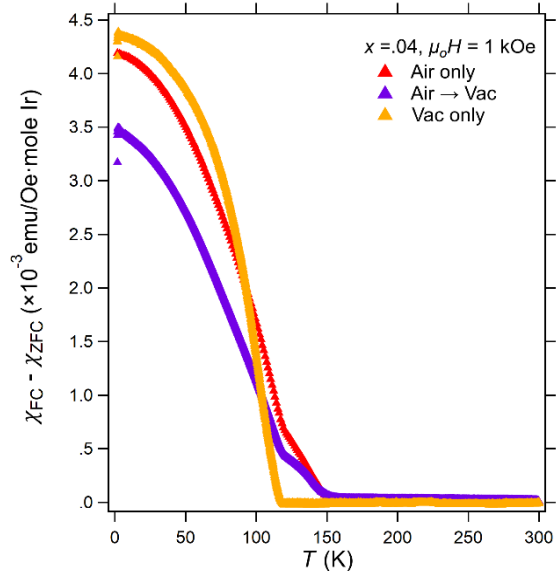

**Figure A:** Comparison of the FC – ZFC magnetic susceptibility for $x = .04$ samples synthesized using different procedures. The "Air only" curve was synthesized completely in air (the first procedure in the **Methods** section). The "Air → Vac" curve is the same sample, following the addition of excess $IrO_2$ and vacuum sintering. The "Vac only" curve is a new sample, synthesized with the complete vacuum sintering

synthesized under high-pressure [19]. In air sintered samples, the strength of the irreversibility at $T'$ trades off with $T_{AFM}$ with Ca-doping until $T'$ eventually dominates the magnetism and $T_{AFM}$ becomes hard to define. However, no upturn in the Ir XMCD (for the air sintered samples) was seen in the region $T_{AFM} < T < T'$. No corresponding signature of $T'$ is seen in the charge transport (**figure B**) which is qualitatively similar to what was seen in the vacuum sintered samples, albeit with a more gradual decrease in $T_{MIT}$. The air sintered parent sample also has a much lower $1/RRR$ value than the vacuum sintered one ( $\approx 490$ and 2500, respectively). The change in $1/RRR$ of the parent sample likely reflects an improved stoichiometry of the Ir sublattice [32], particularly since vacuum sintering reduces the loss of $IrO_2$ by the reaction $IrO_2 + O_2 (g) \rightarrow IrO_3 (g)$ [23].

Analogous behavior was reported in a study on polycrystalline $(Y_{1-x}Ca_x)_2Ir_2O_7$ synthesized by a similar solid-state method [21]. In that case, a secondary transition in $\chi(T)$ develops at $T'$ = 190 K and grows with Ca-doping to dominate the feature at 160 K which corresponds to the $T_{MIT}$ in the $x = 0$ sample. This synthesis dependent phenomenon may therefore be generic in carrier-doped pyrochlore iridates. While the source of this additional feature is currently unresolved, the elimination of $T'$ via vacuum sintering implies the mechanism is related to defects on the Ir sublattice.

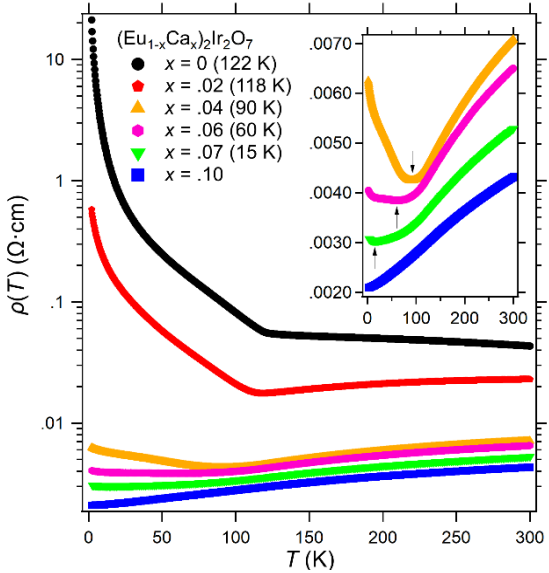

**Figure B:** Resistivity measurements on pellets of air-sintered samples. All data were taken on warming from 2 K to 300 K. The values in the parenthesis are $T_{MIT}$ determined as described in the main text. Inset: closer view of the higher $x$ samples. Black arrows indicate the minima in $\rho(T)$.

## Acknowledgements

This work was supported by ARO Award No. W911NF-16-1-0361 (S.D.W., Z.P., E.Z.). We thank Julian L. Schmehr, Lianyang Dong and Ram Seshadri for helpful discussions. The research reported here made use of the MRL Shared Experimental Facilities, supported by the MRSEC Program of the NSF under Award No. DMR 1720256, a member of the NSF-funded Materials Research Facilities Network (www.mrfn.org). G. L. gratefully acknowledges support for this work from the National Science Foundation (NSF) through DMR 1904980. Additional support was provided by Bates College internal funding (G.L., S.B., S.H). Use of the Advanced Photon Source at Argonne National Laboratory was supported by the U.S. Department of Energy, Office of Science, under Contract No. DE-AC02-06CH11357.